\documentclass{aims}
\usepackage{amsmath}
  \usepackage{paralist}
  \usepackage{graphics} 
  \usepackage{epsfig} 
\usepackage{graphicx}  \usepackage{epstopdf}
 \usepackage[colorlinks=true]{hyperref}
\hypersetup{urlcolor=blue, citecolor=red}

\def\currentvolume{X}
 \def\currentissue{X}
  \def\currentyear{200X}
   \def\currentmonth{XX}
    \def\ppages{X--XX}
     \def\DOI{10.3934/xx.xx.xx.xx}

\title[Non-Holonomic Constraints in Klein-Gordon Lattices]
{Non-holonomic constraints and their impact on discretizations
of Klein-Gordon lattice dynamical models}

\author[P.G. Kevrekidis and V. Putkaradze and Z. Rapti]{}


 \email{kevrekid@math.umass.edu}
 \email{putkarad@ualberta.ca}
 \email{zrapti@math.uiuc.edu}

\begin{document}
\maketitle

\centerline{\scshape Panayotis G.\ Kevrekidis}
\medskip
{\footnotesize
 \centerline{Department of Mathematics and Statistics}
  \centerline{University of Massachusetts}
 \centerline{Amherst, MA 01003-4515, USA}

}
\medskip

\centerline{\scshape Vakhtang Putkaradze}
\medskip
{\footnotesize
 \centerline{Department of Mathematics and 
Statistics, University of Alberta }
 \centerline{Edmonton, Alberta, CA}

}

\medskip

\centerline{\scshape Zoi Rapti}
\medskip
{\footnotesize
 \centerline{Department of Mathematics, 
University of Illinois at Urbana-Champaign} 
 \centerline{Urbana, Illinois 61801-2975, USA}

}

\begin{abstract}
We explore a new type of discretizations of lattice dynamical models of the Klein-Gordon
type relevant to the {existence and long-term mobility }of nonlinear waves. 
The discretization is based on non-holonomic constraints and is shown to retrieve
the ``proper'' continuum limit of the model. Such discretizations
are useful in {exactly preserving a discrete analogue of the
momentum. It is also shown that for generic initial data, the momentum and energy 
conservation laws cannot be achieved concurrently}. Finally, direct numerical simulations illustrate
that {our models yield considerably higher mobility of strongly nonlinear 
solutions than the
well-known ``standard'' discretizations, even in the case of highly
discrete systems when the  coupling between the adjacent
nodes is weak. Thus, our approach is better suited for  cases where an accurate description of 
mobility for nonlinear traveling waves is important}.
\end{abstract}

\section{Introduction}

Discrete solitons such as kinks and breathers exist in many
physical systems, including Josephson juctions, electrical circuits, and
granular crystals \cite{ourbook, dodd, boechler}.
In the models that describe these systems, the mobility of soliton solutions, and 
specifically kinks -- which will be the focus of the present work--, is an important
and well-studied property \cite{peyrard_krusk, pgk_dmitriev}. For instance,
kinks in extemely discrete systems need to overcome the 
celebrated Peierls-Nabarro barrier, 
while they undergo energy radiation which leads to their trapping in the lattice 
\cite{kivshar1993}. More recent work revealed the existence of kinks that do not experience
the action of the Peierls-Nabarro potential \cite{dmitriev2005}.

For a class of Klein-Gordon discretizations \cite{pgk_dmitriev} it has been shown that
energy and linear momentum cannot be both conserved. This is problematic in cases 
when the objective is the discretization of the continuum version of equation, which
conserves both momentum and energy. Going beyond the properties of isolated kinks, the above 
result has consequences for the studies of soliton collisions in Klein-Gordon lattices. 
In general, the mobility of kinks in discrete manifestations of the Klein-Gordon 
equation may vary from that in the continuous version as well as among different
discretizations.

In this work, we present a class of discretizations based on a non-holonomic constraint 
that preserve the momentum and alleviate the problem of decreased mobility in kink solutions.  
Specifically, as we enforce the conservation of momentum it in turn implies an
affine constraint in kink velocities, which is of the non-holonomic type. 
The extra term added in the discretization due to the
non-holonomic constraint, vanishes in the continuum limit, hence
demonstrating that the discretized version of the equation converges to its
continuous counterpart.  Our numerical results validate our theoretical considerations and
we present multiple instances where the additional constraint that we impose increases the 
mobility of the kinks.

We shall note that for generalized non-holonomic constraints that are not linear in velocities, 
one has to be careful in applying the variational principle. This was demonstrated in, for example, 
\cite{marle, cendra} for the case of stabilizing servomechanisms, where an alternative formulation through Poisson  structures has been derived. The discrete equivalent of a conservation law for continuous system (\ref{eqn1}) below, enforced by the constraint, cannot be readily interpreted as a consequence of a continuous symmetry through a Noether theorem. Thus, applications of alternative methods of non-holonomic mechanics \cite{ArKoNe2006}, in addition to the Lagrange-d'Alembert's principle developed here, are of particular interest and will be considered in future work.  

This article is structured as follows. In section \ref{sec:model} we describe the 
Klein-Gordon model that will be studied and present the theoretical framework of
our method. In section \ref{sec:numer} we present numerical results that 
corroborate our theoretical predictions. Finally, in section \ref{sec:discuss}
we discuss our findings and offer insight for future directions.
    

\section{Model Theoretical Setup and Proposed Discretizations}
\label{sec:model}

The models that we will consider herein will involve discretizations
of continuum partial differential equations {of the nonlinear wave type}: 
\begin{eqnarray}
u_{tt}= u_{xx} - V'(u). 
\label{eqn1}
\end{eqnarray}
There are
numerous key model examples of this {class of equations,} such as the 
sine-Gordon~\cite{ourbook} or the $\phi^4$ model arising from classical 
field theory and high-energy physics~\cite{dodd}, as well as
variants thereof.

The corresponding discrete models are of considerable
interest in their own right as models of coupled torsion 
pendula, dislocation dynamics, 
Josephson junctions, electrical circuits, among
numerous other settings~\cite{ourbook,dodd,remois,peyrard}.
The generic discrete analogue of these models
{is derived by using a standard lattice 
discetization of the second space derivative as follows: }
\begin{eqnarray}
\ddot{u}_n=\epsilon (u_{n+1} + u_{n-1} -2 u_n) - V'(u_n)
\label{eqn2}
\end{eqnarray}
The field is defined at the integer nodes $n$ 
and $\epsilon$ denotes the coupling constant
between adjacent sites, 
{which could be interpreted as $\epsilon=1/dx^2$, with $dx$ being 
the spacing between the adjacent sites (see below). } 
 
Note that  lattice model (\ref{eqn2}) is Hamiltonian with 
\begin{eqnarray}
H = \sum_n {\frac{1}{2} \dot u_n^2 +} \frac{{\epsilon} }{2} \left(u_{n+1} - u_n\right)^2 + V(u_n) 
\equiv \sum_n {\mathcal H}_n,
\label{eqn4}
\end{eqnarray}
where ${\mathcal H}_n$ is the energy density.

In that context, an alternative
limit, namely $\epsilon=0$, the so-called anti-continuum limit 
becomes particularly important and provides a valuable tool for 
assessing the potential stationary or time-periodic states of the
model and their respective spectral stability 
properties~\cite{macaub,aubry97,pgk_review}.

One of the ``pathologies'' of such discrete
models is that as the {coupling parameter $\epsilon$} decreases,
the mobility of the nonlinear waves accordingly weakens~\cite{peyrard_krusk}.
{The decreased mobility stems, to a considerable
degree,  from the fact that the momentum 
$P= \int u_t u_x dx$ is no longer a conservation law in the
discrete case of Eq.~(\ref{eqn2}), even though it is 
in the corresponding continuum limit of $\epsilon \rightarrow
\infty$. 
This curious fact } has spurred a considerable volume of recent work
(for a partial summary, see~\cite{pelinov}) focused on providing
alternative discretizations that would enable higher mobility of
the waves, even at small $\epsilon$.
Our aim in the present work is to propose a novel class of
such discretizations, based upon the imposition of
a non-holonomic constraint  
providing an exact conservation of the discrete  momentum.

Indeed, the continuum equation (\ref{eqn1}) provides a conservation of the momentum quantity 
$P_c=\int u_t u_x \mbox{d} x$. On the other hand, after defining the analogue of discrete momentum for  
(\ref{eqn2}) as 
\begin{eqnarray}
P = \sum \dot{u}_n (u_{n+1} - u_{n-1})
\label{eqn5}
\end{eqnarray}
a quick check shows that such discrete momentum is not conserved \cite{physd}. 
The goal of this paper is to \emph{enforce} the conservation of (\ref{eqn5}) using the methods of 
non-holonomic mechanics. The equation 
(\ref{eqn2}) can be obtained as the Euler-Lagrange equation with the Lagrangian 
\begin{equation} 
L=\sum \frac{1}{2} \dot u_n^2  - \left( \frac{ \epsilon}{2} (u_{n+1}-u_n )^2 + V(u_n) \right) \, .  
\label{Lagr} 
\end{equation}
through the critical action principle $\delta\int L \mbox{d} t =0$ on arbitrary variations $\delta u_n(t)$ with 
appropriately vanishing boundary conditions at the end time points. 
Next, enforcing the conservation law  (\ref{eqn5}) leads to an affine constraint in velocities of the type 
\begin{equation} 
\sum_n  a_n( \mathbf{u} ) \dot u_n = b(\mathbf{u})   \qquad {\rm with } \qquad 
a_n(\mathbf{u}) = u_{n+1}-u_{n-1}\,, \,  \quad b(\mathbf{u})=P   
\label{affine_constr}
\end{equation} 
using the notation $\mathbf{u}=(u_1, u_2, \ldots , u_n)$.
Since this constraint involves velocities in a non-trivial way, \emph{i.e.}, that constraint cannot be integrated 
to contain coordinates only, the constraint is non-holonomic, and linear in velocities. The usual way of dealing with this 
type of constraints is through the Lagrange-d'Alembert's principle \cite{bloch2009}. 
For the constraints of the type (\ref{affine_constr}), the variations $\delta u_n$ are no longer arbitrary, 
but must satisfy the linear relations  
\begin{equation}
\sum_n a_n( \mathbf{u} ) \delta u_n= \sum_n \left(u_{n+1} - u_{n-1} \right)  \delta u_n =0 \, . 
\label{Lagr_dalembert} 
\end{equation} 
Notice that we do not enforce (\ref{affine_constr}) directly, as this would correspond 
to the so-called \emph{vakonomic} approach, relevant for \emph{e.g.} the control 
theory but normally not leading to correct expressions for mechanical systems, as described in \cite{ArKo2007}. 
On the contrary, Lagrange-d'Alembert's approach (\ref{Lagr_dalembert}) 
was consistently  shown to be physically appropriate. For more in-depth discussion 
of various approaches to non-holonomic 
systems, we refer the reader to 
\cite{bloch2009,ArKo2007}. 

If $r$ is the Largange multiplier enforcing the constraint  (\ref{Lagr_dalembert}), 
the critical action principle coupled with the constraint gives 
\[ 
\int_{t_0}^{t_1} \left[  \frac{d}{dt} \frac{\partial L}{\partial \dot u_n } - 
\epsilon \left(u_{n+1} + u_{n-1} -2 u_n \right)+ V'(u_n) - 
r\left(u_{n+1} - u_{n-1} \right)  \right] \delta u_n  =0\, ,\] 
provided $\delta u_n$ vanish at $t=t_{0,1}$ so no boundary integration terms appear in the equations.  
Since $\delta u_n$ is arbitrary, the equations of motion conserving momentum are 
\begin{eqnarray}
\ddot{u}_n=\epsilon \left(u_{n+1} + u_{n-1} -2 u_n \right) - V'(u_n) + 
r (u_{n+1} - u_{n-1}) \, . 
\label{eqn3}
\end{eqnarray}
Note the new term proportional to $r$ and enforcing the constraint on the right 
hand side of (\ref{eqn3}),  as compared to (\ref{eqn2}). 
The physical meaning of $r$ is the (generalized) constraint force. In order to find $r$, we 
compute the derivative $dP/dt=0$ and set it equal to zero.  We find that
while the terms associated with $\epsilon$ vanish telescopically~\cite{physd},
the nonlinear term does not. This leads to the explicit condition for
$r$ as 
\begin{eqnarray}
r=\frac{\sum_n V'(u_n) \left(u_{n+1} - u_{n-1} \right)}{\sum_n \left(u_{n+1} - u_{n-1}\right)^2}
\label{eqn6}
\end{eqnarray}
Making this selection for $r$,   
 multiplying both sides of Eq.~(\ref{eqn3}) by $\dot{u}_n$,
and summing, we observe that  the derivative of the (former) Hamiltonian
of Eq.~(\ref{eqn4}) is not zero anymore, but is given by the following condition: 
\begin{eqnarray}
\frac{d H}{dt}=r P.
\label{eqn7}
\end{eqnarray}
Importantly, this identity suggests that for {\it generic}
initial data, such that $P \neq 0$, the Hamiltonian
will not be conserved. This is due to the constraint (\ref{affine_constr}) 
being non-homogeneous in velocities for $P \neq 0$. 
It is relevant to point out that
such an ``incompatibility'' of the momentum and energy
conservation laws for other broad classes of discretizations
of the Klein-Gordon dynamical lattices have also 
been reported previously \cite{pgk_dmitriev}. 

An additional comment about this 
{non-holonomic discretization is that  the extra term 
introduced by the non-holonomic constraint vanishes  at the continuum
limit $\epsilon \rightarrow \infty$}, ensuring that Eq.~(\ref{eqn3}) converges to the proper
continuum limit of Eq.~(\ref{eqn1}). To see this, {we consider
the continuum analogue of the new $r (u_{n+1} - u_{n-1})$ term, which is }
 $u_x (\int V'(u) u_x dx/ \int u_x^2 dx$). 
The numerator of the term in the parenthesis can be integrated
to yield $V(u_2)-V(u_1)$ where $u_2$ and $u_1$ are, respectively
the asymptotic values of the field at $\pm \infty$. Assuming
that we examine the dynamics of homoclinic orbit or of a
heteroclinic orbit in a symmetric double well (as is typically
the case in the examples of interest mentioned above), this quantity
vanishes identically and the discretization reverts to its proper
continuum limit. We will return to this point and corroborate
it through our numerical computations below.
A final remark about the { nature of our discretization procedure, different from all other
schemes that we are aware of,  is that our method is  inherently  nonlocal since the 
computation of  $r$ in (\ref{eqn6}) entails a 
summation over the entire lattice. 

\section{Numerical Computations}
\label{sec:numer}

To provide a comparison between the properties of the different 
discretizations, we examine the dynamics of Eq.~(\ref{eqn2})
vs. that of Eq.~(\ref{eqn3}). We use an initial condition
in the form of a prototypical nonlinear excitation such as the
kink. In the continuum limit, the latter stationary solution
(which can be boosted to any sub-luminal speed via
Lorentz transformation) is of the form $u(x)=\tanh(x)$.  This is
the case for the $\phi^4$ model with a potential $V(u)=(1/2)
(u^2-1)^2$. 
For the discrete system, we have used the initial condition 
$u_n(0)=\tanh(x)=\tanh(n \times dx)$, where 
$dx$ is used to signal the spatial discretization step (see also below), approximating the continuous solution on a discrete set of points. 
Furthermore, Eqs~. (\ref{eqn2}) and (\ref{eqn3}) are
integrated using a fourth-order Runge-Kutta scheme with a temporal discretization step  of the typical size
$dt=0.00025$. 

In addition, we enforce 
an initial momentum of $P_{fix}=0.5$ {via the definition of Eq.~(\ref{eqn5})}. 
The simplest way that we envisioned
for achieving this was to assume all sites as having vanishing
velocity except for the central one of the chain, say $n_0$, and
then using $\dot{u}_{n_0}=P_{fix}/(u_{n_0+1}-u_{n_0-1})$.
This prescription provides the given amount of momentum to the
central portion of the kink and we expect that it should 
set the kink in motion.

This is exactly what we observe in Fig.~\ref{fig_1vp}, which
is close to the continuum limit of $\epsilon=1/dx^2 \rightarrow \infty$.
{Here, as usual $dx$ is effectively the corresponding spacing between the lattice
sites, leading  as a finite-difference three-point 
approximation to the second derivative. } In particular, in this
figure $dx=1/4$ or $\epsilon=16$ is used. The top left panel illustrates
the traveling kink of the standard discretization of Eq.~(\ref{eqn2}).
The proximity of the model to its continuum 
analogue coupled with the imparting of momentum via the initial
condition leads the kink to travel through the chain with an
apparently constant speed (evident in the constant slope of
the two contour plots of the top panel, the left one illustrating
the energy density of Eq.~(\ref{eqn4}) and the right one
the actual field $u_n(t)$). Given that the model of Eq.~(\ref{eqn2})
is Hamiltonian, our explicit fourth-order time integration
(Runge-Kutta) scheme conserves the initial energy (of O$(1)$)
up to a few parts in $10^{-14}$ verifying its conservation
for all practical purposes. On the other hand, the momentum
fluctuates in the third decimal digit indicating its non-preservation
by the scheme; instead, it is only {approximately} conserved because of the
proximity to the continuum limit where the continuum analogue
of Eq.~(\ref{eqn5}) can be shown to be preserved by Eq.~(\ref{eqn1}).

\begin{figure}
\begin{tabular}{cc}
 \includegraphics[width=6cm]{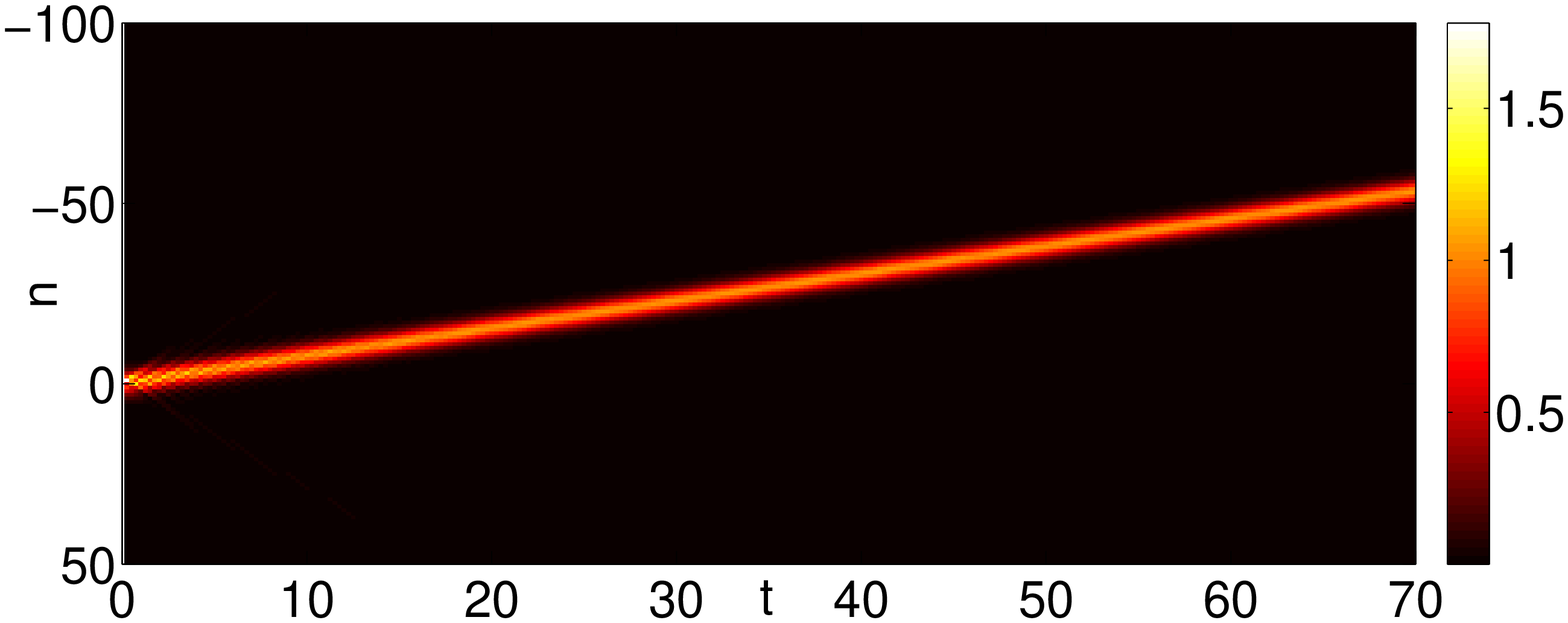} &
 \includegraphics[width=6cm]{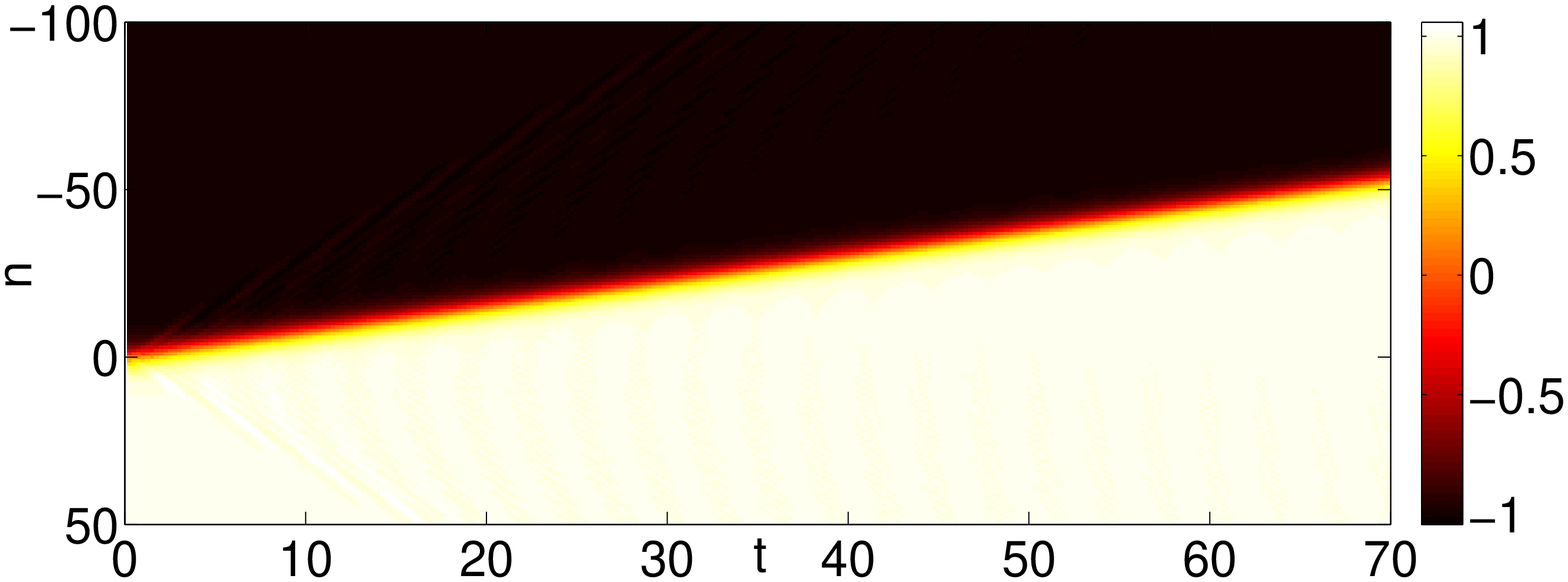} \\
\includegraphics[width=6cm]{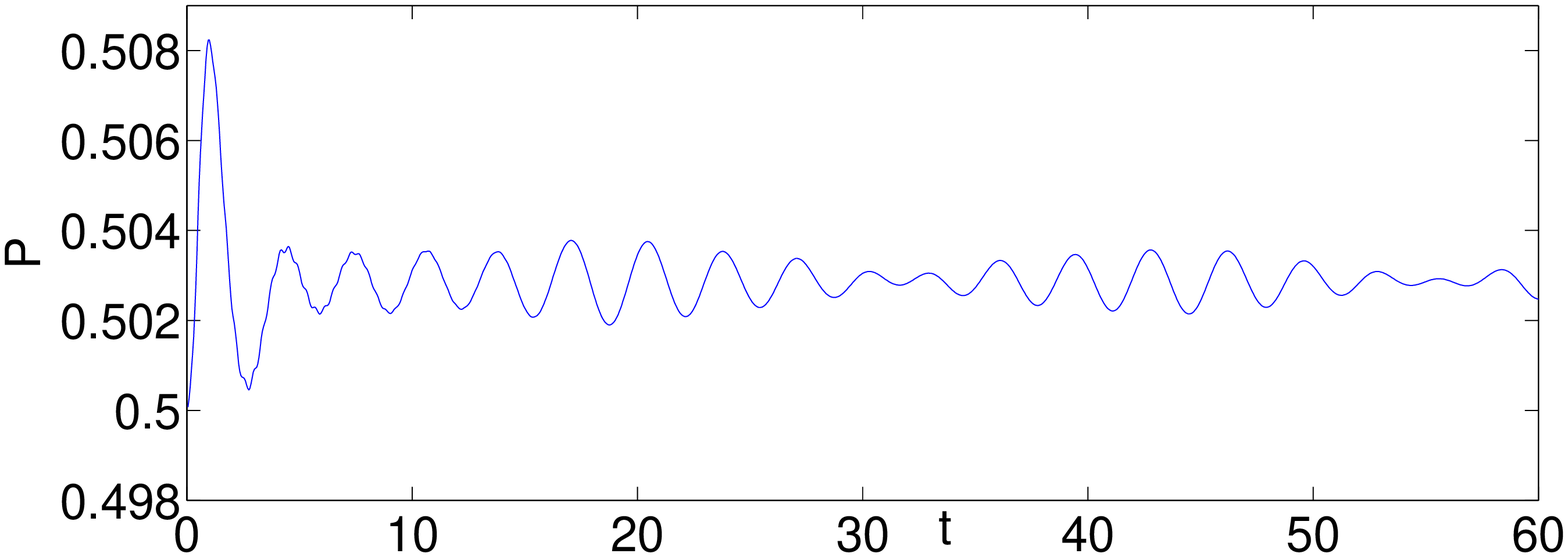} &
 \includegraphics[width=6cm]{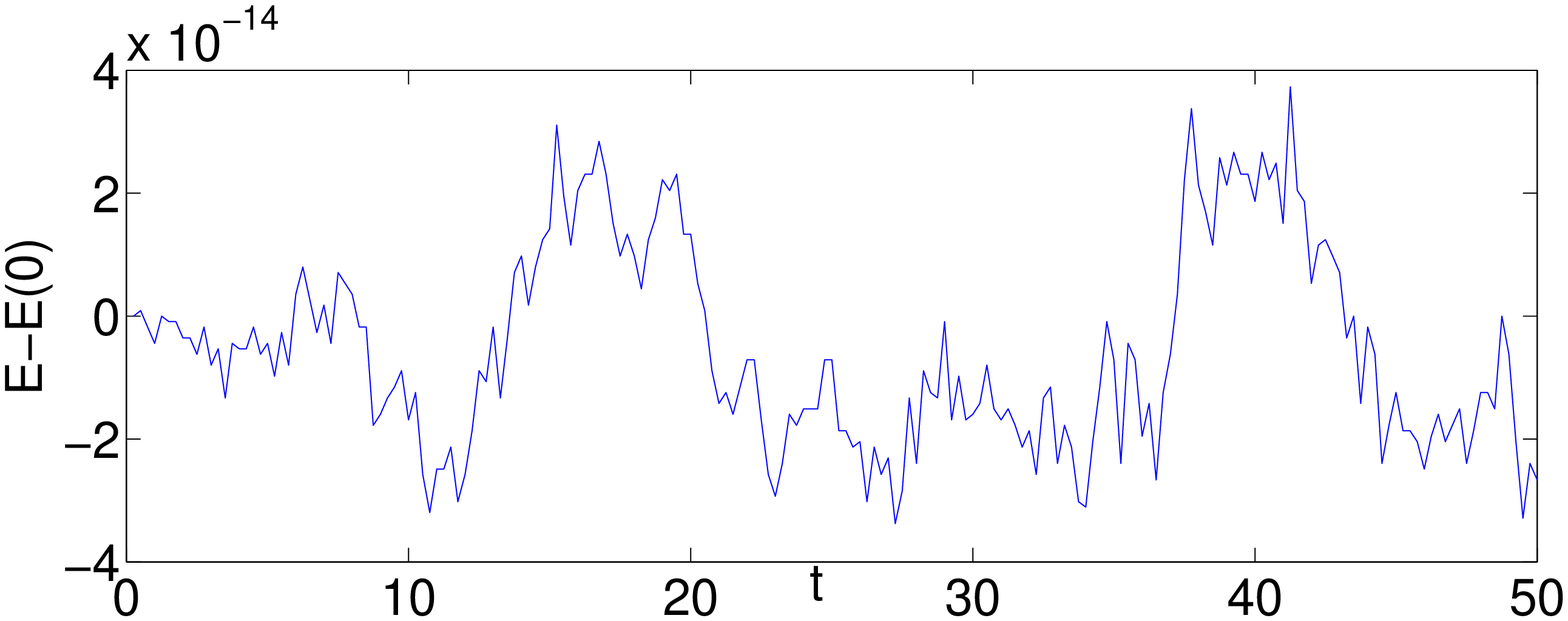} \\
\end{tabular}
\caption{Evolutionary dynamics of the ``standard'' discrete model
of Eq.~(\ref{eqn2}) with $dx=1/4$, hence $\epsilon=1/dx^2=16$ for
$V(u)=\frac{1}{2} (u^2-1)^2$ i.e., the $\phi^4$ model. 
The top left panel represents the space-time ($n-t$) contour plot
of the energy density (see Eq.~(\ref{eqn4})), while the top right
represents the corresponding contour of the field $u_n(t)$.
The bottom left illustrates the ``near-conservation'' for this
(approaching the continuum) case for the linear momentum
of Eq.~(\ref{eqn5}), while the bottom right illustrates that the
model is {energy preserving}  up to fluctuations of size of $10^{-14}$ 
due to numerical roundoff errors.}
\label{fig_1vp}
\end{figure}

On the other hand, we can see that the fluctuations in this
quantity ($P$) are far more dramatic in the considerably
more discrete case 
$\epsilon=1$, corresponding to $dx=1$}, shown on  Fig.~\ref{fig_2vp}. 
There, {the fluctuations in $P$ are 
of O$(1)$  while the energy is still conserved with the same precision as before. 
In addition, the motion of the kink is fundamentally different,
featuring a jerky back and forth trajectory between a few sites around
the origin. Thus, in this case,  the kink appears trapped due to the
weak inter-site coupling which does not allow the momentum to 
translate itself into mobility as it would in the corresponding
continuum limit.

\begin{figure}
\begin{tabular}{cc}
 \includegraphics[height=2.5cm, width=6cm]{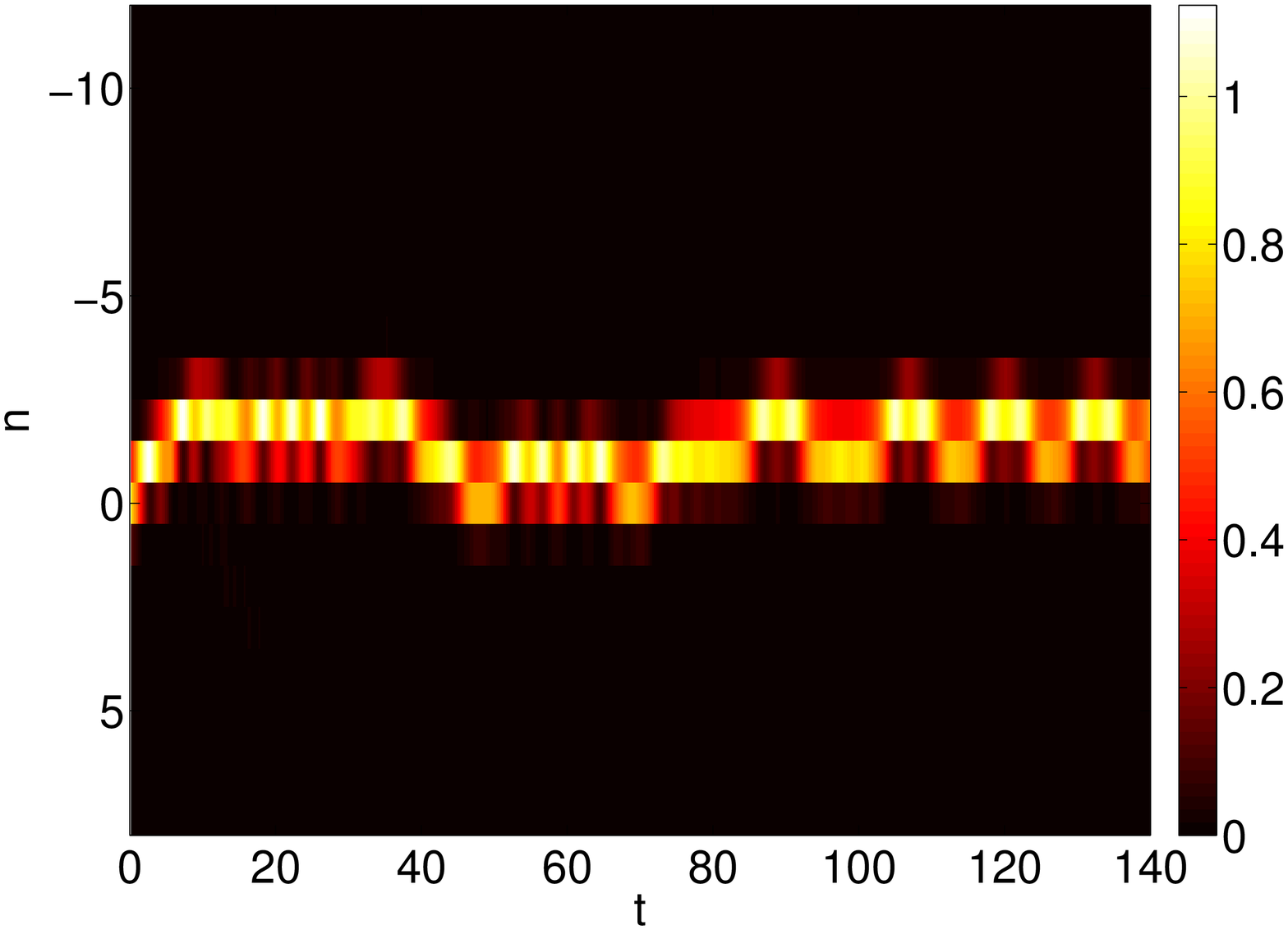} &
 \includegraphics[width=6cm]{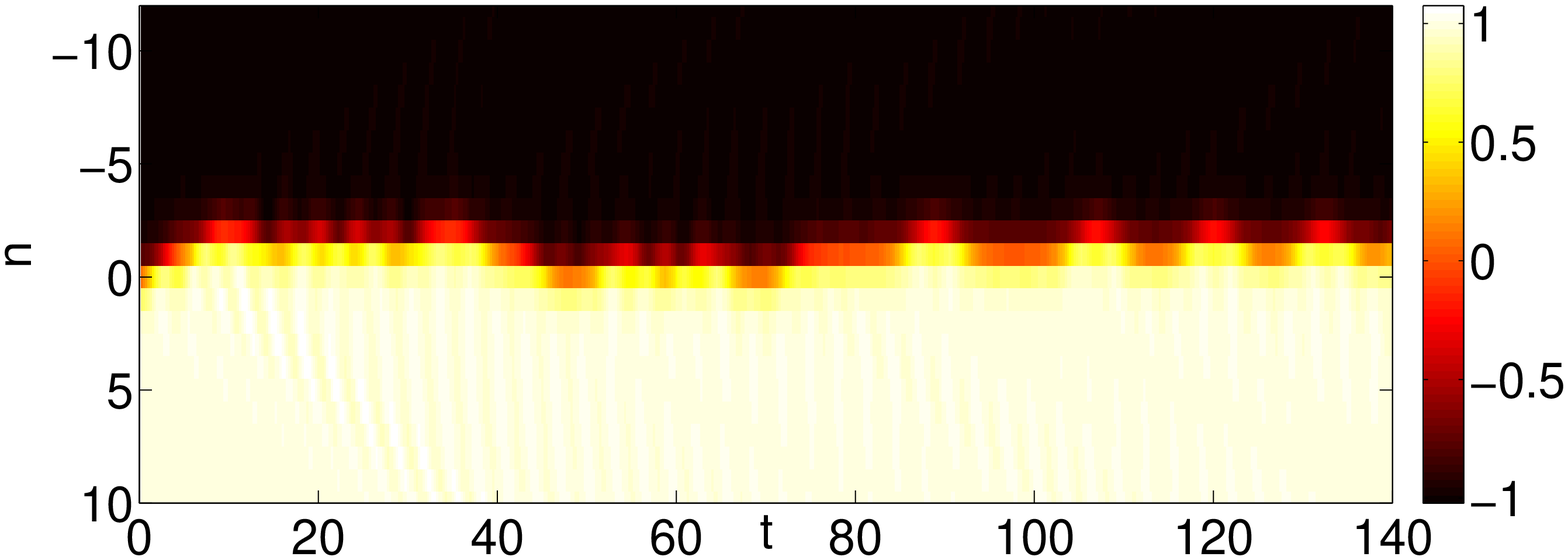} \\
\includegraphics[width=6cm]{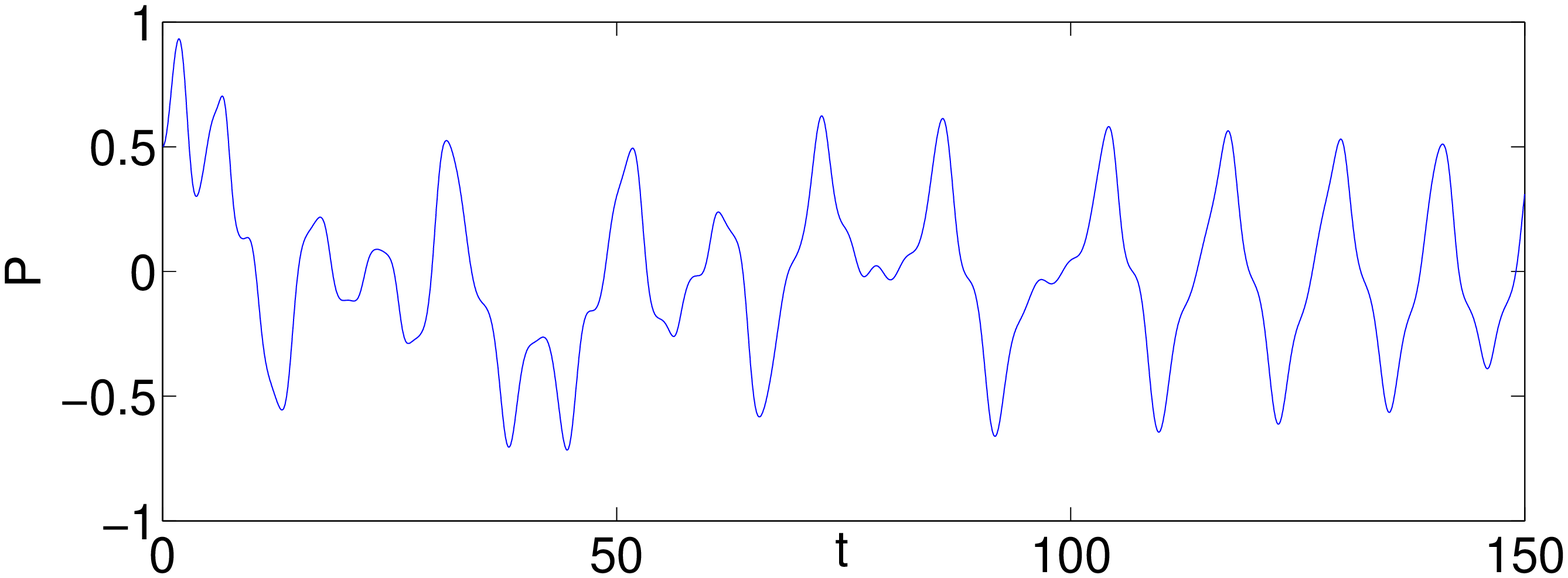} &
 \includegraphics[width=6cm]{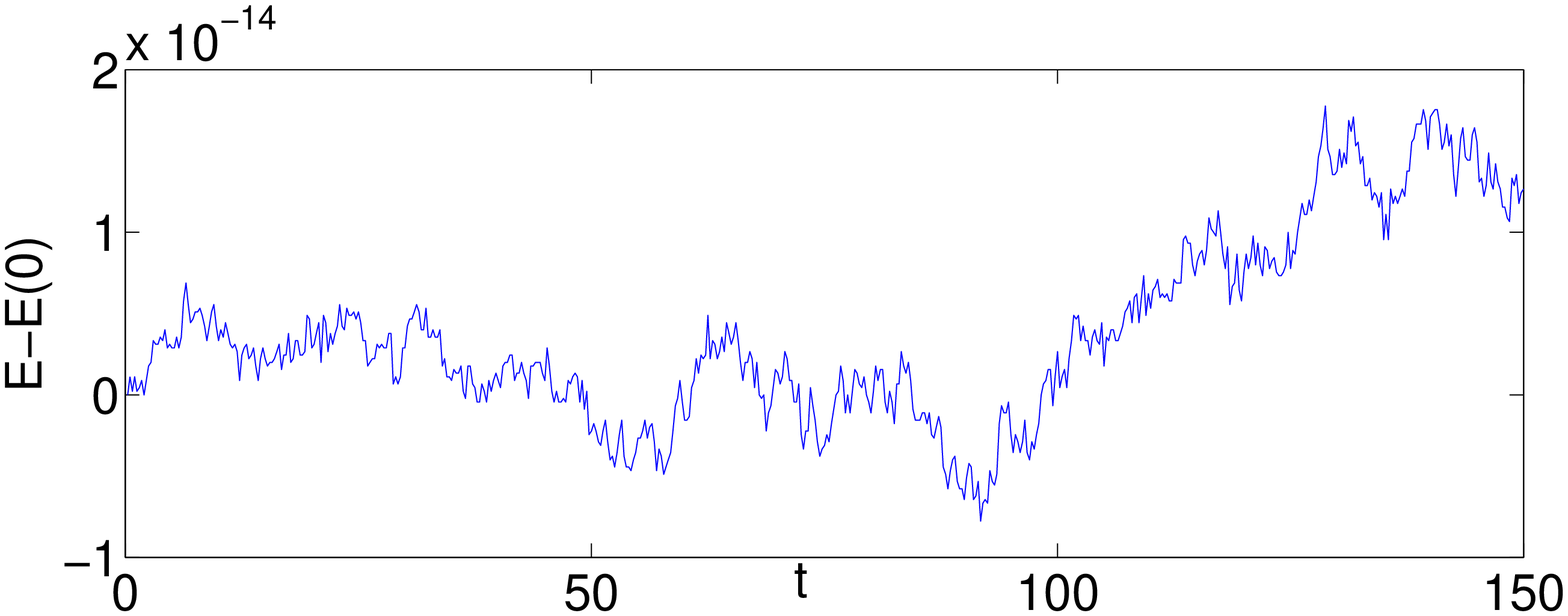} \\
\end{tabular}
\caption{Same as the previous plot, only now we consider a highly discrete case 
 $dx=\epsilon=1$. Here, we observe the detrimental effect of the
energy-conserving discretization with regards to the mobility of the
kink initial condition. The strong discreteness/weak coupling precludes
the kink from moving through the lattice and the initial momentum (which
as can be seen is far from being conserved now) is expended in a jerky
motion with a range of a few discrete sites (trapping of the kink).}
\label{fig_2vp}
\end{figure}

We now turn to the corresponding dynamics of the newly proposed model
of the form of Eq.~(\ref{eqn3}). The case of $dx=1/4$ is again
shown in Fig.~\ref{fig_3vp}. The fundamental difference in this case
is that the momentum $P$ of Eq.~(\ref{eqn5}) 
in the bottom left panel is conserved to a similar
accuracy as the energy in the previous model, as expected
by our construction. On the contrary, the energy of the bottom
right panel varies at the third significant digit remaining
roughly constant solely due to the proximity to the continuum
limit, where the analogue of Eq.~(\ref{eqn4}) is conserved
for Eq.~(\ref{eqn1}).

\begin{figure}
\begin{tabular}{cc}
 \includegraphics[width=6cm]{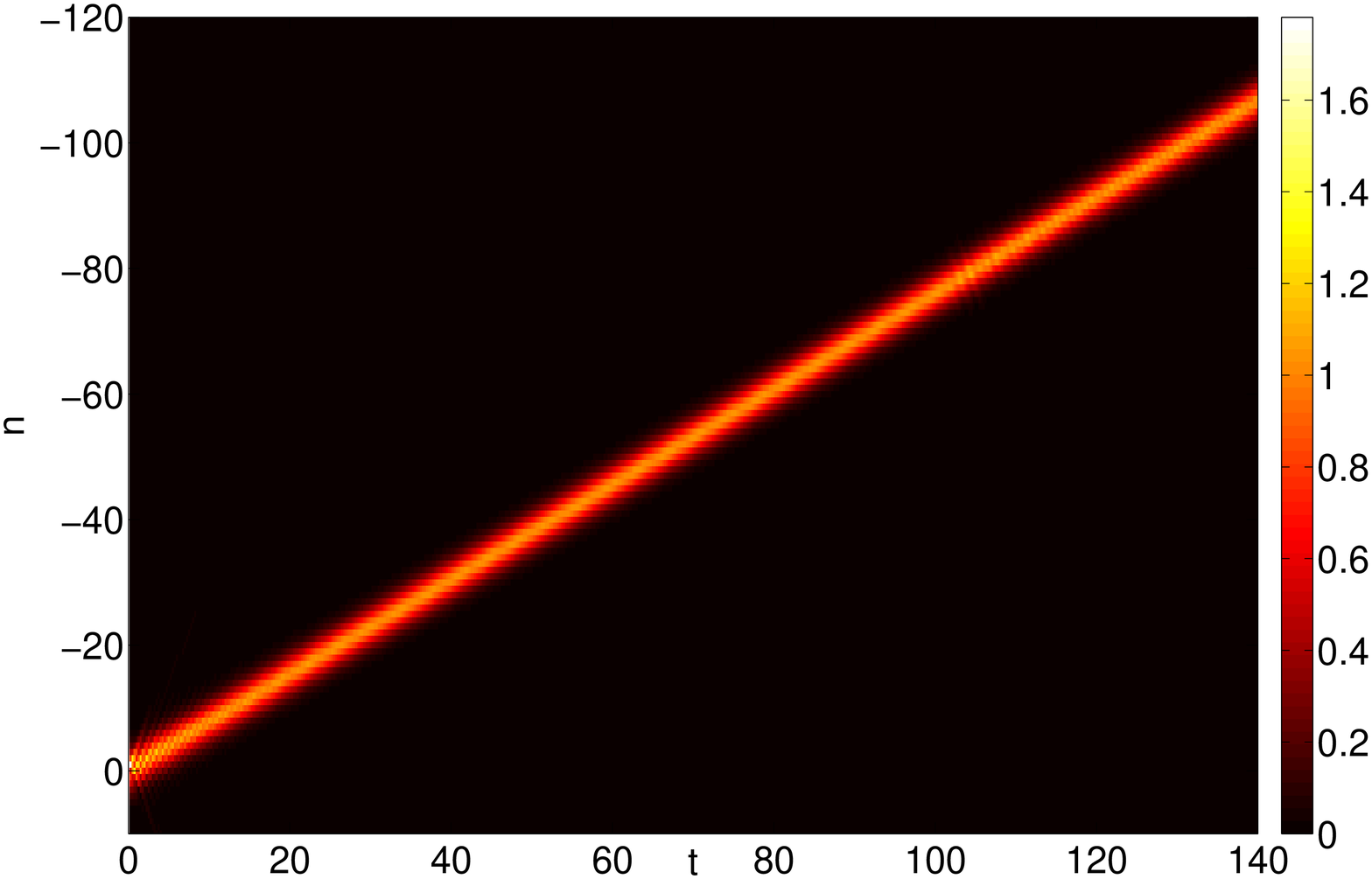} &
 \includegraphics[width=6cm]{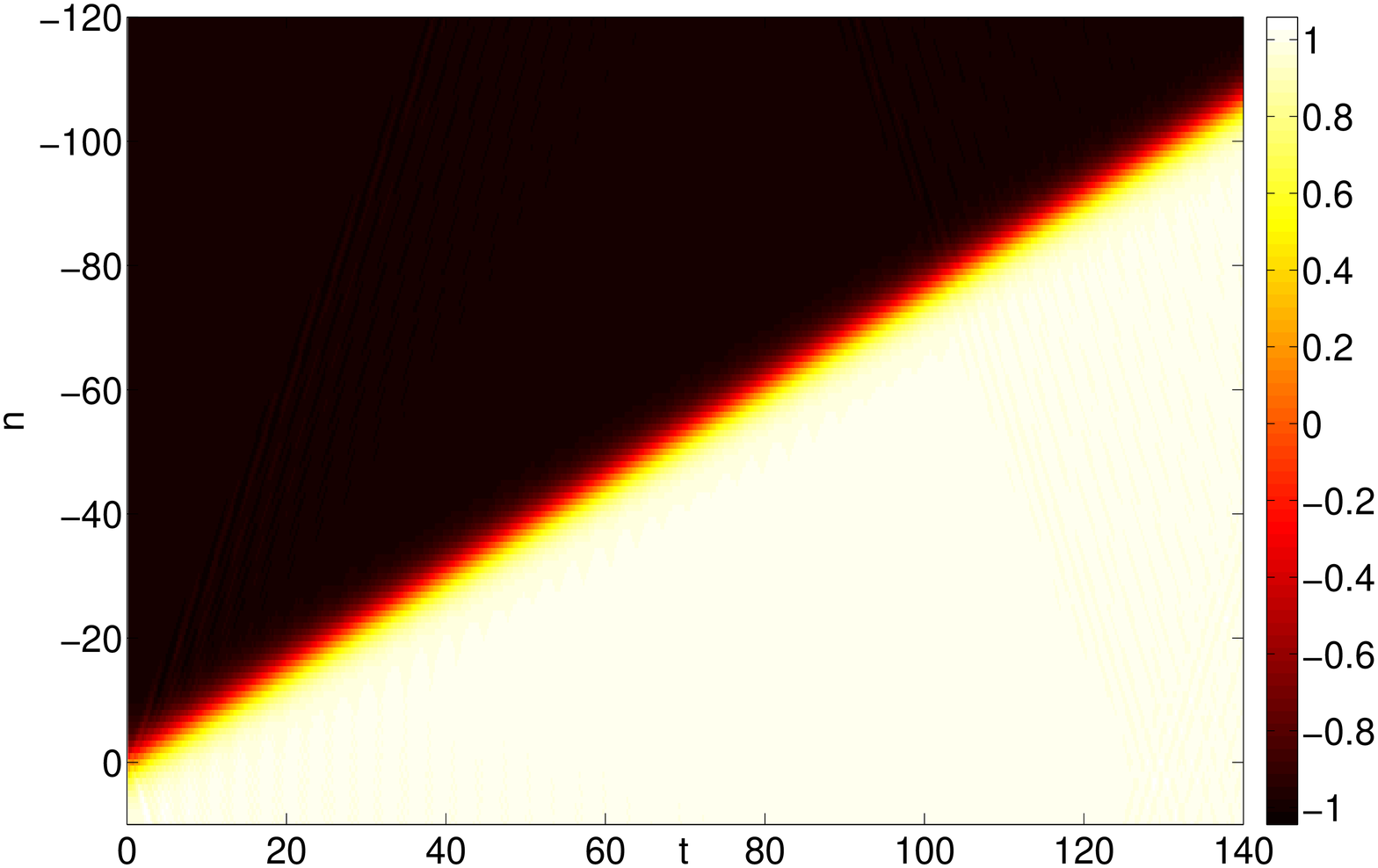} \\
\includegraphics[width=6cm]{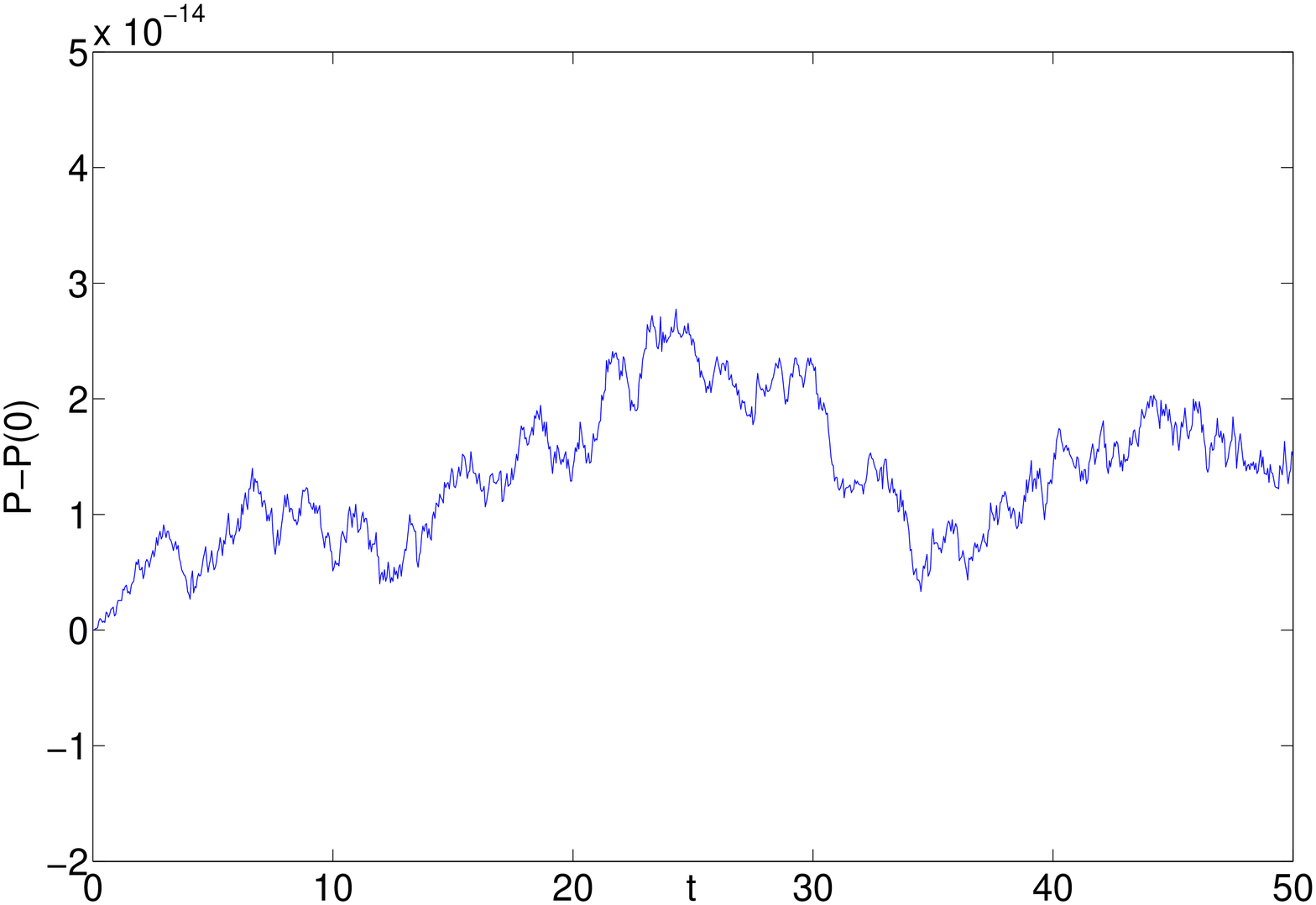} &
 \includegraphics[width=6cm]{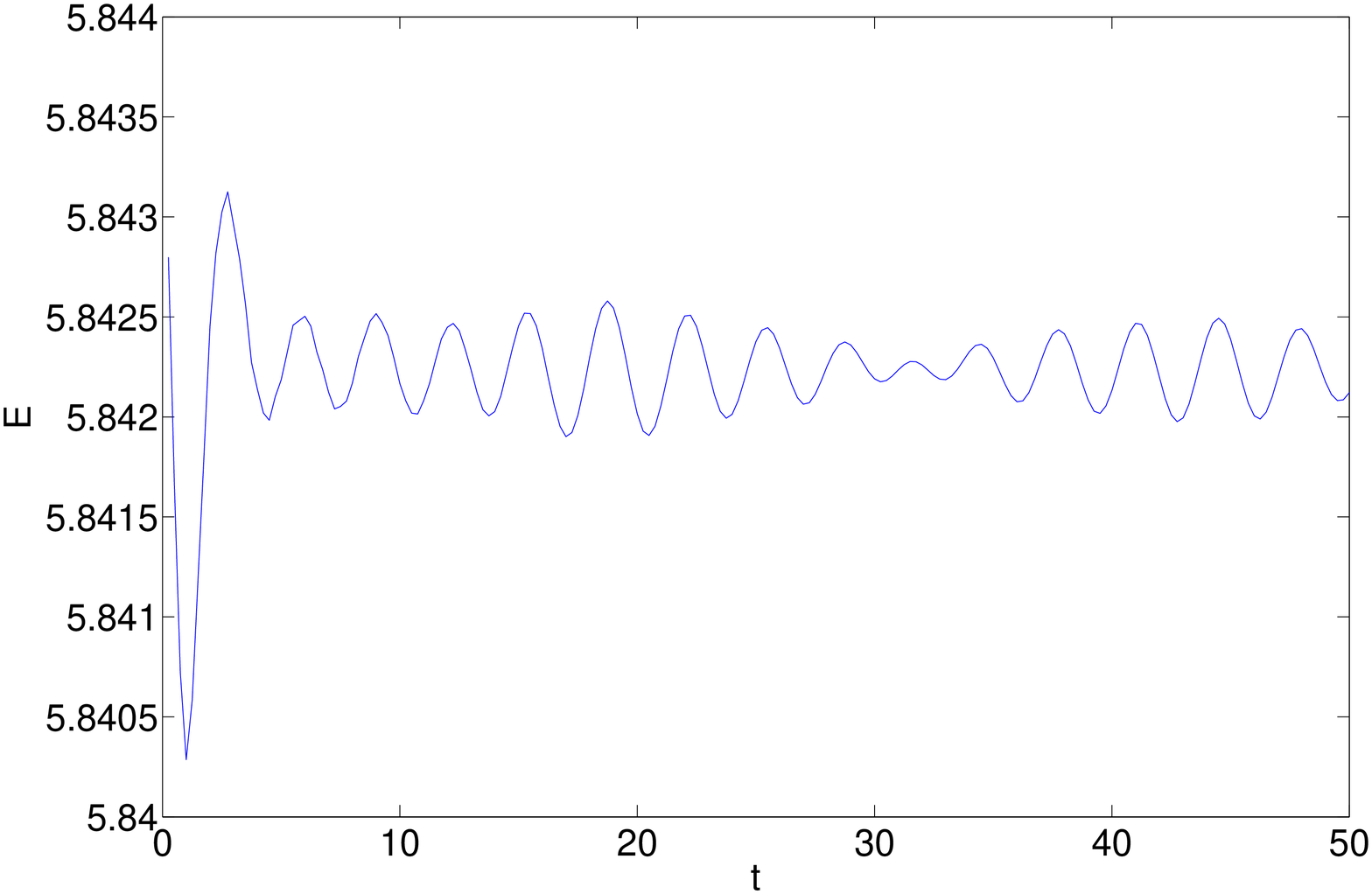} \\
\end{tabular}
\caption{Same as Fig.~\ref{fig_1vp} but now for the momentum
conserving discretization of Eq.~(\ref{eqn3}). This is the near
continuum case of $dx=0.25$ which sustains motion of the kink
with constant speed. The momentum of the bottom left is
{ conserved up to numerical accuracy}, while the energy of Eq.~(\ref{eqn4}) 
in the bottom right is only approximately conserved, due to the
proximity to the energy conserving continuum limit.}
\label{fig_3vp}
\end{figure}

The main advantage of our discretization method 
 is evident in Fig.~\ref{fig_4vp}.
More specifically, as the example of  $dx=\epsilon=1$ shown in Fig.~\ref{fig_3vp} illustrates, 
the discretization of Eq.~(\ref{eqn3}) presented here can accurately describe traveling waves in 
the case of strong coupling. 
Fig.~\ref{fig_3vp} shows that the momentum
preserved up to numerical accuracy, as expected,
while the energy fluctuations become more significant, revealing
the absence of energy conservation already in the second significant
digit. It should also be mentioned that we have also separately verified Eq.~(\ref{eqn7}) numerically, 
 although we do not present it here.  Hence, our principal
numerical conclusion is that the models of Eq.~(\ref{eqn3})
ensure a higher mobility of their corresponding coherent
structures compared with the standard discretization
of Eq.~(\ref{eqn2}). This comes at the price
of the incorporation of the nonlocal effect through
the summation that ensures the conservation of the linear
momentum. 

\begin{figure}
\begin{tabular}{cc}
 \includegraphics[width=6cm]{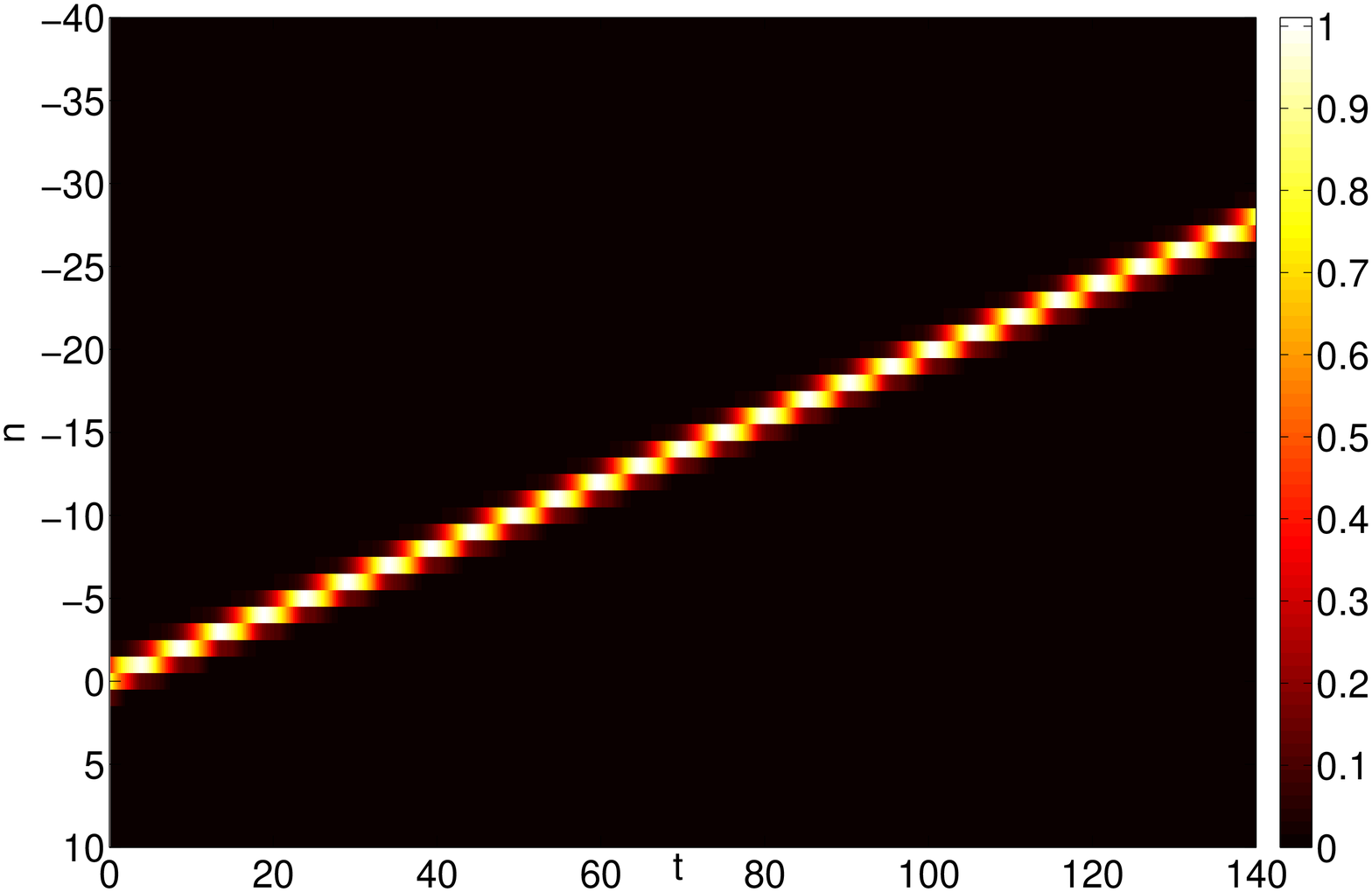} &
 \includegraphics[width=6cm]{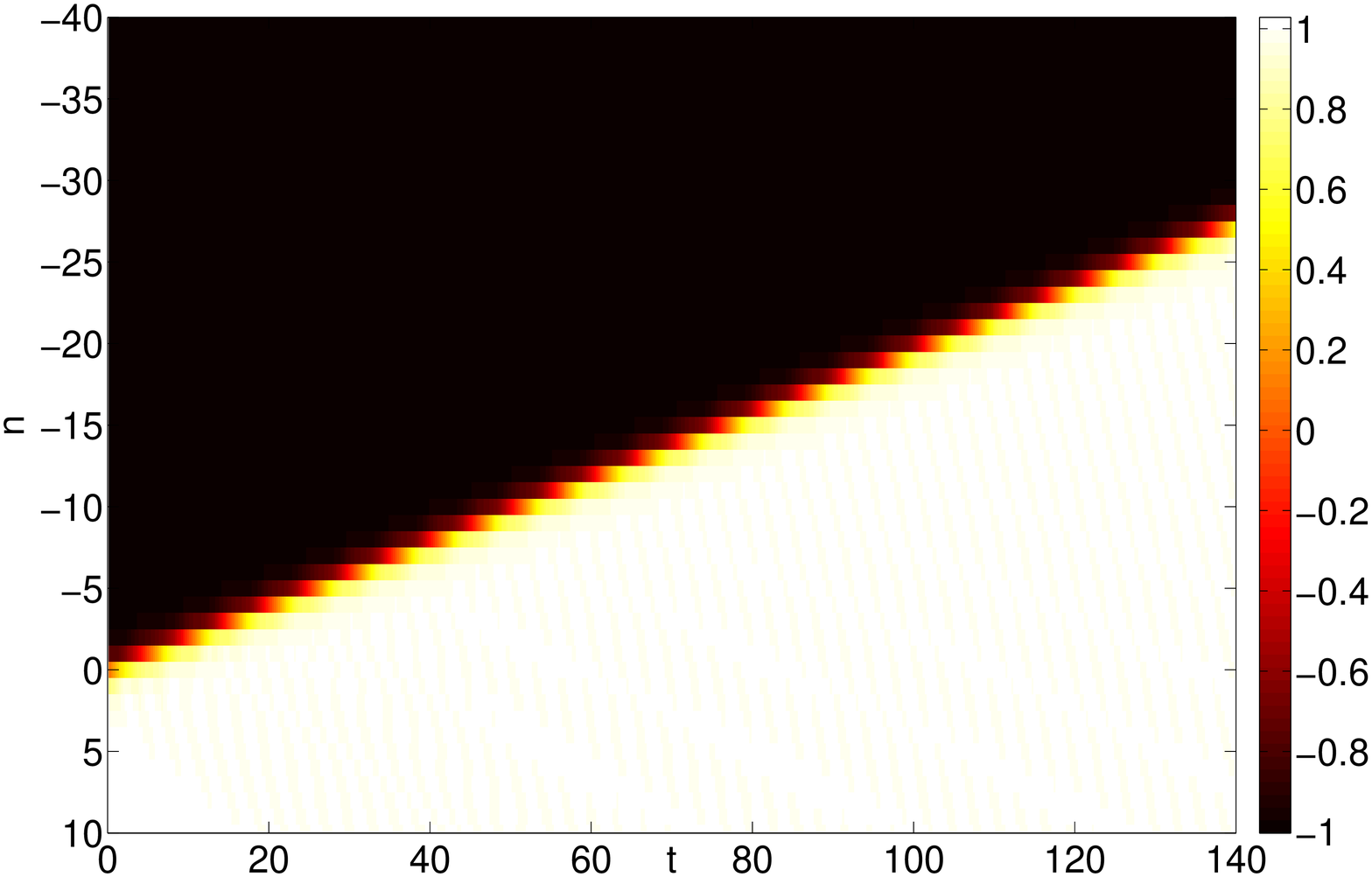} \\
\includegraphics[width=6cm]{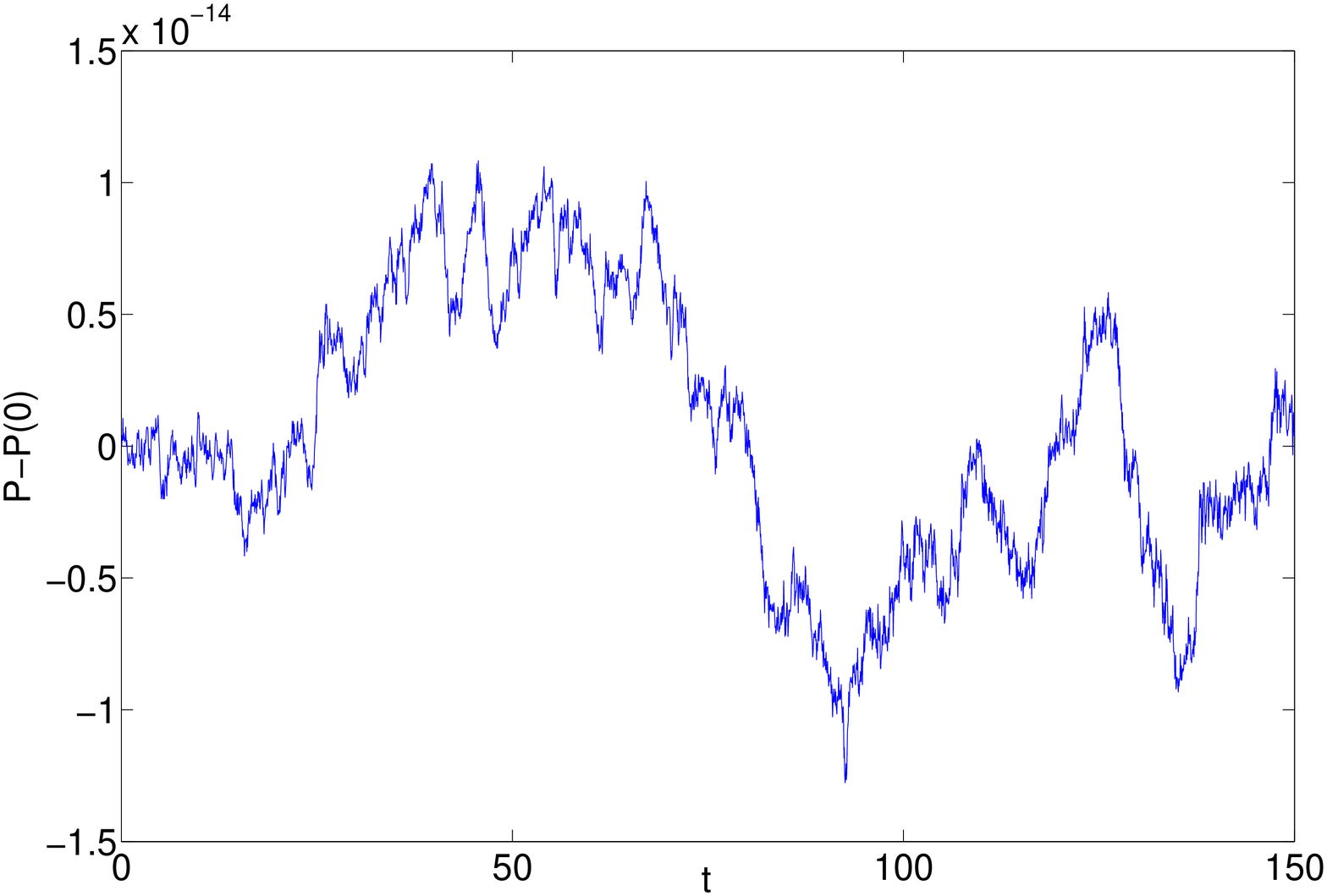} &
 \includegraphics[width=6cm]{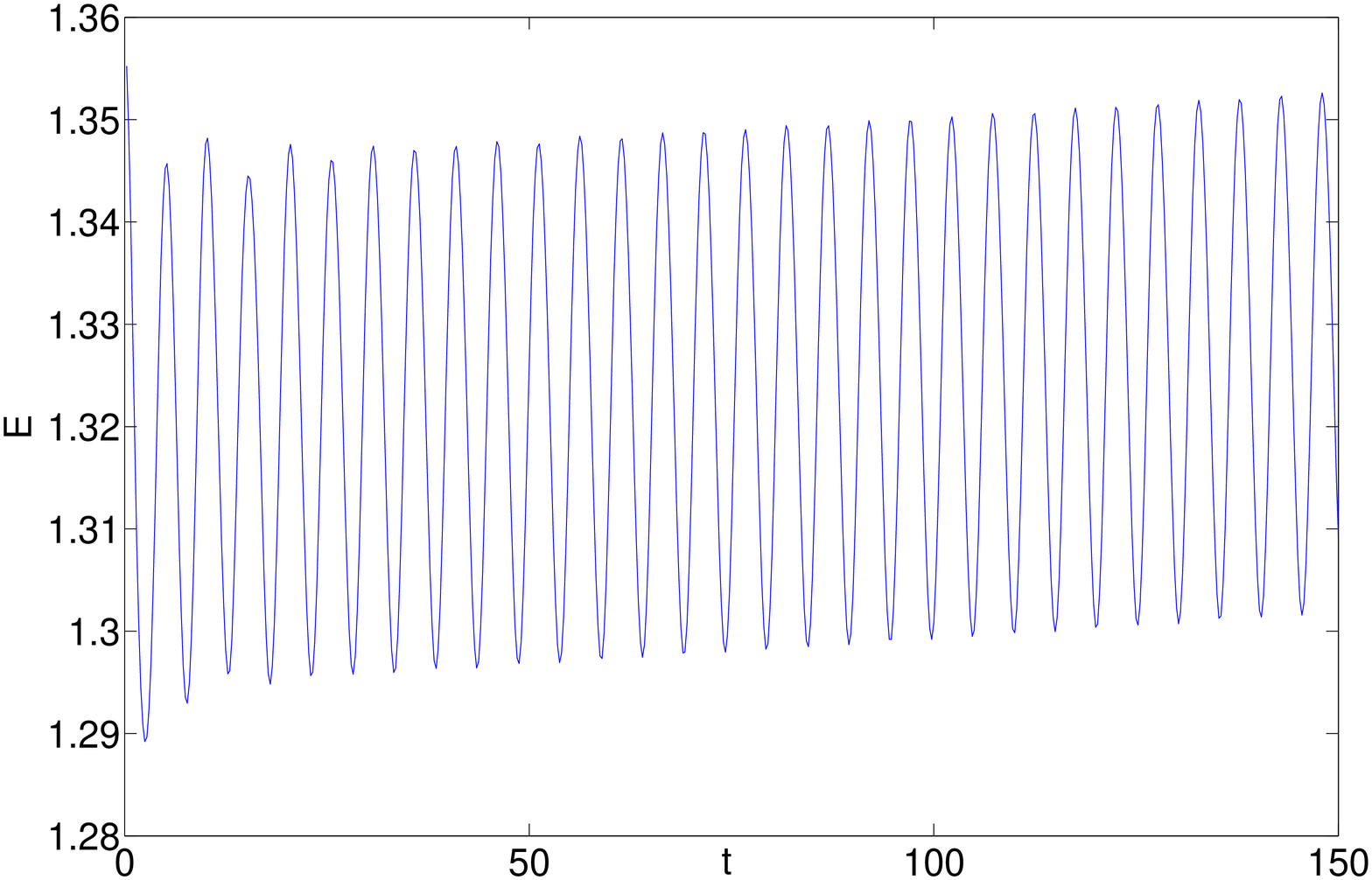} \\
\end{tabular}
\caption{Same as Fig.~\ref{fig_4vp} but now for $\epsilon=1$ (corresponding to $dx=1$). Notice
that despite the high discreteness of the model and the weakness
of the coupling, the mobility of the state is well preserved.}
\label{fig_4vp}
\end{figure}

Lastly, we have confirmed that the relevant discretizations are
indeed properly converging to the continuum limit as $dx \rightarrow
0$. A typical example of this approach is shown in Fig.~\ref{fig_5vp}.
In particular, the figure showcases the evolution of the quantity
$r(t)$, which constitutes the prefactor of
the newly proposed nonlocal term arising
through the non-holonomic constraint.
It can be seen that as we go from the weakly-coupled strongly
discrete case of the left panel (for $dx=1$) to the right
panel, closer to continuum case of $dx=0.25$, the value of
$r(t)$ decreases by more than an order of magnitude

\begin{figure}
\begin{tabular}{cc}
 \includegraphics[width=6cm]{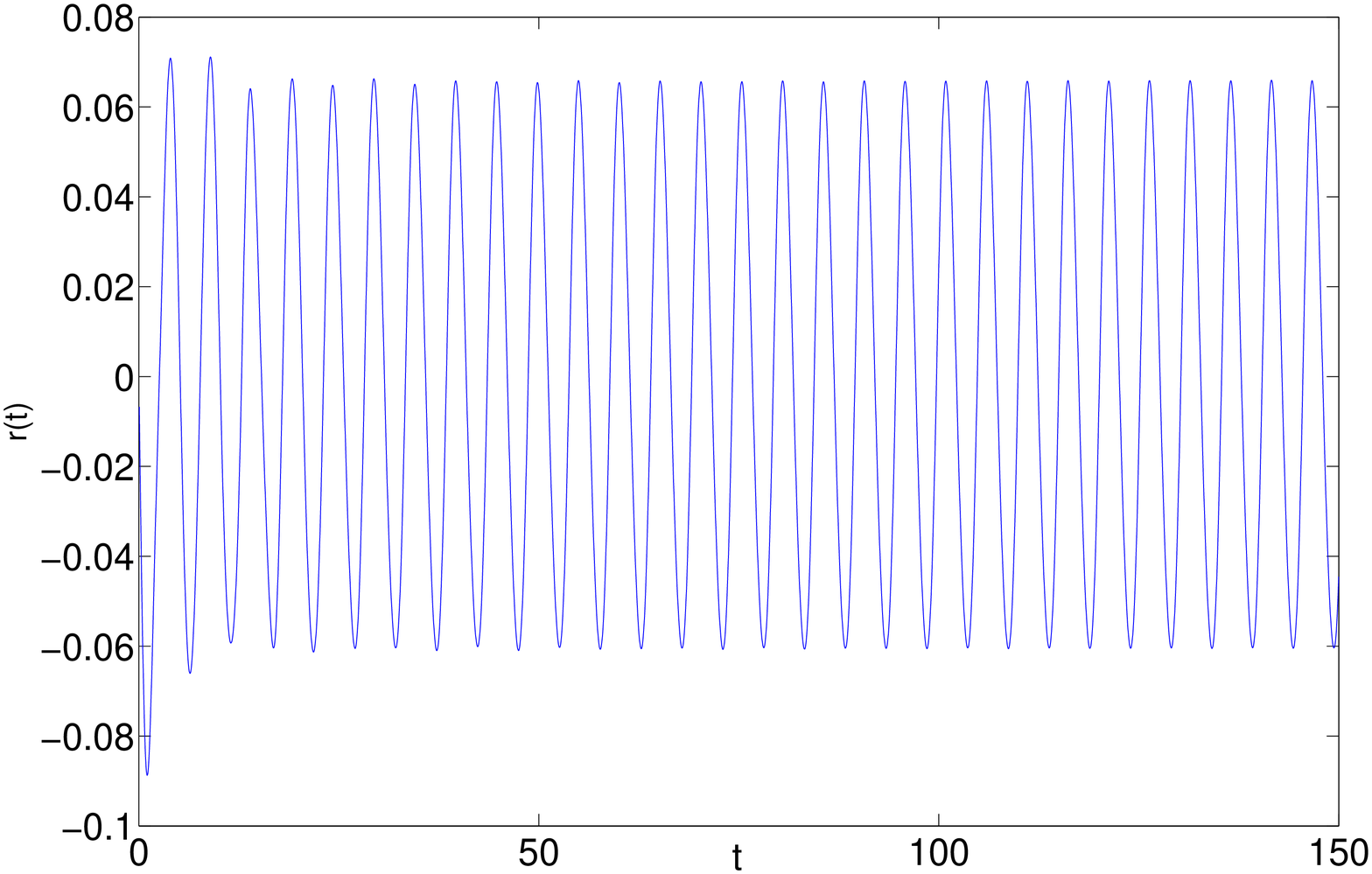} &
 \includegraphics[width=6cm]{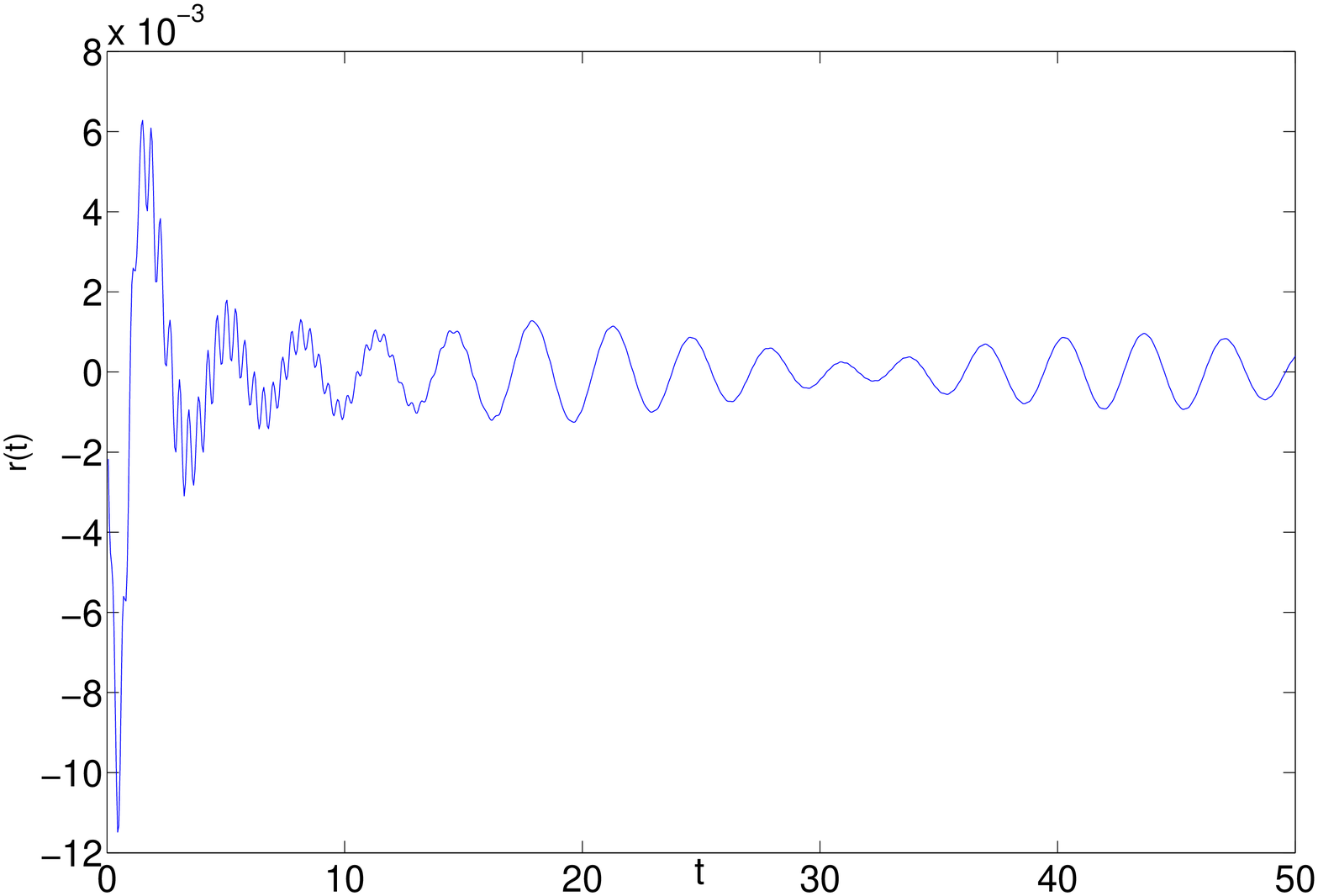} \\
\end{tabular}
\caption{Corroboration of the convergence of the discrete
model of Eq.~(\ref{eqn3}) to the continuum
limit of Eq.~(\ref{eqn1}). The time evolution of the 
quantity $r(t)$ is shown, in the left panel for $dx=1$
and in the right for $dx=0.25$. It can be seen that the quantity
assumes significantly smaller values in the latter case in 
comparison to the former, and indeed this trend continues for 
decreasing values of $dx$, with $r \rightarrow 0$, as $dx \rightarrow 0$.}
\label{fig_5vp}
\end{figure}

\section{Conclusions and  Future Challenges}
\label{sec:discuss}

In the present work, we have revisited the theme of discretizations
of continuum nonlinear partial differential equations of the wave
type. We have focused on the example of Klein-Gordon equations and
offered a novel class of schemes which can be summarized as follows: 
 (a) they can be straightforwardly 
designed to conserve linear momentum; (b) when they do so, they
do not {in general conserve the energy, }  (c) they revert to the
continuum PDE in the appropriate limit of the lattice spacing
tending to zero; and (d) they are based on the imposition of
a non-holonomic constraint and thus are fundamentally different
than previously proposed schemes due to their nonlocal
nature. 
We have argued that these models have intriguing characteristics,
such as the ability to impart high mobility to localized
coherent states, a feature corroborated via numerical computations
in the case of kinks of the $\phi^4$ model. This feature is
preserved even in the regime of high discreteness/weak
coupling where the standard discretizations under the same
initial data produce kinks that are trapped within the
Peierls-Nabarro energy barrier and cannot move through the
chain.

This study opens an interesting new direction for the investigation
of the relevance of such schemes from a numerical perspective and
also for a more detailed understanding of their properties.
While in the present initial communication we have focused on the
dynamics of kinks, it would be relevant to explore corresponding
structural features for discrete breather type solutions~\cite{gorbach}.
Furthermore, while in the present work we have focused on the
one-dimensional {domains only}, it would be particularly interesting to
consider generalizing such models in higher dimensions. The latter
have progressively become a theme of considerable attention
over the past few years~\cite{minzoni,caputo} and hence it would
again be particularly relevant to compare the standard higher
dimensional discretizations, with the ones produced by the
multi-dimensional generalization of the non-holonomic constraint
concepts presented herein. Such studies are presently in progress
and will be reported in future publications.

\section*{Acknowledgements}  The first author is supported by NSF-DMS-1312856,
the US-AFOSR under grant FA9550-12-10332, from FP7-People under grant IRSES-606096 and the 
Binational (US-Israel) Science Foundation through grant
2010239. The second author is supported 
by NSERC Discovery grant and the University of Alberta. 
The third author is supported by NSF-EFRI-1024772. We also acknowledge fruitful discussions with Prof. D. V. Zenkov.

\end{document}